\documentclass[twocolumn,tighten,times,twocolappendixm]{aastex631}
\usepackage{amsmath}
\usepackage{multirow}
\usepackage{scalerel}
\usepackage{graphicx} 
\usepackage{orcidlink}
\usepackage{hyperref}

\newcommand{\HI}{\text{H\,\textsc{i}}} 
\newcommand{\solM}{\ensuremath{M_\odot}}
\newcommand{\MHI}{\ensuremath{M_{\mathrm{\HI}}}}
\newcommand{\VHI}{\ensuremath{V_{\mathrm{\HI}}}}
\newcommand{\Mstar}{\ensuremath{M_{*}}}
\newcommand{\Wfifty}{\ensuremath{W_{50}}}
\newcommand{\kms}{\ensuremath{\rm{km\,s^{-1}}}}
\newcommand{\xHI}{\ensuremath{x_{\mathrm{\HI}}}}
\newcommand{\Hnot}{\ensuremath{H_{\mathrm{0}}}}
\newcommand{\Omegam}{\ensuremath{\Omega_{m}}}

\newcommand{\Mpcdex}{\ensuremath{\mathrm{Mpc^{-3}\,dex^{-1}}}}
\def\arcsec{$''$}
\def\arcmin{$'$}

\begin{document}

\title{CHORD \HI-Galaxy Survey Forecasts: Searching for nearby dark galaxies and high redshift giants}

\author{Akanksha Bij\,\orcidlink{0000-0001-7505-5223}}
\affiliation{Department of Physics, Engineering Physics and Astronomy, Queen’s University, Kingston, ON K7L 3N6, Canada} 

\author{Kristine Spekkens\,\orcidlink{0000-0002-0956-7949}}
\affiliation{Department of Physics, Engineering Physics and Astronomy, Queen’s University, Kingston, ON K7L 3N6, Canada}

\author{Hans S. Hopkins\,\orcidlink{0009-0002-1199-8876}}
\affiliation{Perimeter Institute for Theoretical Physics, 31 Caroline Street N, Waterloo, ON N25 2YL, Canada}
\affiliation{Waterloo Center for Astrophysics, University of Waterloo, Waterloo, ON N2L 3G1, Canada}

\author{Michael G. Jones\,\orcidlink{0000-0002-5434-4904}}
\affiliation{IPAC, Mail Code 100-22, Caltech, 1200 E. California Boulevard, Pasadena, CA 91125, USA}

\author{Arnab Chakraborty\,\orcidlink{0000-0002-7758-9859}}
\affiliation{Department of Physics, McGill University, 3600 Rue University, Montreal, QC H3A 2T8, Canada}
\affiliation{Trottier Space Institute, 3550 Rue University, Montreal, QC H3A 2A7, Canada}

\author{Simon Foreman\,\orcidlink{0000-0002-0190-2271}}
\affiliation{Department of Physics, Arizona State University, Tempe, AZ 85287, USA}

\author{Alex S. Hill\,\orcidlink{0000-0001-7301-5666}}
\affiliation{Department of Math, Physics, \& Statistics, University of British Columbia, Okanagan Campus, Kelowna, BC V1V 1V7, Canada}
\affiliation{Dominion Radio Astrophysical Observatory, Herzberg Research Centre for Astronomy and Astrophysics, National Research Council, Penticton, BC V0H1K0, Canada}

\author{Dustin Lang\,\orcidlink{0000-0002-1172-0754}}
\affiliation{Perimeter Institute for Theoretical Physics, 31 Caroline Street N, Waterloo, ON N25 2YL, Canada}
\affiliation{Waterloo Center for Astrophysics, University of Waterloo, Waterloo, ON N2L 3G1, Canada}

\author{Adrian Liu\,\orcidlink{0000-0001-6876-0928}}
\affiliation{Department of Physics, McGill University, 3600 Rue University, Montreal, QC H3A 2T8, Canada}
\affiliation{Trottier Space Institute, 3550 Rue University, Montreal, QC H3A 2A7, Canada}

\correspondingauthor{Akanksha Bij}
\email{a.bij@queensu.ca}

\begin{abstract}
Population studies of gas-rich galaxies across the full range of Neutral Hydrogen (\HI) masses that galaxies are known to exhibit (\mbox{$10^5 \lesssim \MHI \lesssim 10^{11} \solM$}) remain limited by the need to conduct high-sensitivity, wide-band surveys across significant sky areas. The Canadian Hydrogen Observatory and Radio-transient Detector (CHORD) is a next-generation radio telescope that will significantly expand the census of \HI-galaxies to date from untargeted drift-scan surveys at declinations  $+20^{\circ} < \delta < +80^{\circ}$. We draw survey realizations from a known \HI\ mass function (HIMF) to forecast \HI\ detections in fiducial 1-year and 5-year CHORD surveys. The 5-year survey source counts is expected exceed currently available catalogs by roughly an order of magnitude, notwithstanding the potential impacts of radio frequency interference (RFI) and spectroscopic source confusion that we also estimate. We predict that CHORD will push the low-mass \HI\ galaxy census to $\MHI \sim 10^{5.5}\solM$, over an order of magnitude lower than has been previously achieved. At the high mass end of the HIMF, CHORD is expected to detect $\sim10^{3}$ massive gas-rich giants ($\MHI \gtrsim 10^{10.5}\, \solM$) at $0.3 \lesssim z \lesssim  0.5$, which will explore the evolution of this population relative to local universe estimates. CHORD \HI\ surveys will therefore improve our understanding of the neutral gas reservoirs at the low-mass and high-mass extremes of the galaxy population.
\end{abstract}

\section{Introduction}
Neutral Hydrogen (\HI) extragalactic surveys provide insights on galaxy formation and evolution within the Lambda Cold Dark Matter ($\Lambda$CDM) framework \citep[e.g.\ ][]{HISurveys}. In particular, measurements of the \HI\ mass function (HIMF) probe the atomic gas content of the galaxy population, measuring the statistics of how this gas is accreted and retained across dark matter halos of different masses. Several \HI\ surveys have observed that the local ($z\sim0$) HIMF follows a Schechter distribution \citep{Schechter} with an exponential decline in number density towards higher masses and a power-law increase towards lower masses \citep[e.g.,][]{Rosenberg2002, Springbob2005, Zwaan2005, Martin10, ALFALFA_Jones2018, Ma2025}. However, both the low and high mass extremes of this function are currently poorly constrained, the former due to sensitivity limitations and small survey volumes sampled for the faintest sources in existing surveys, and the latter because of the relative rarity of the most massive sources, leading to minimal source counts in existing surveys. Thus, from an \HI\ perspective, neither the high nor low mass extreme of the galaxy formation process is well understood, and there are few constraints on how the HIMF evolves over cosmic time.

A turnover in the low-mass end of the HIMF is predicted from theoretical models that suggest a redshift-dependent mass threshold ($M_{\rm{halo}}\lesssim 10^{8} \solM$) below which dark matter halos are unable to retain their \HI\ gas due to feedback from the ultraviolet background \citep[e.g.,][]{Efstathiou1992, Cai2014, Finkelstein2019}. Only more massive halos ($\gtrsim 10^{9.5} \solM$) possess gravitational potential wells deep enough to trap gas and ignite star formation \citep{Rees1977}. This mass-dependent suppression of galaxy formation is proposed to reconcile the discrepancy between the large number of low-mass dark matter halos predicted by $\Lambda$CDM and the relatively small number of observed faint galaxies \citep[e.g.,][]{Klypin1999, Moore1999, Sawala2016, Sales2022}. Nevertheless,  direct and definite observations of non-luminous halos are limited \citep[e.g.,][]{Zackrisson2010, Sawala2017, Lu2026}.

In the intermediate mass regime ($M_{\rm{halo}}\sim 10^{8}-10^{9}\solM$), cosmological simulations identify a population of halos that retain their pristine \HI\ gas reservoirs post cosmic reionization, yet remain completely starless or `dark' \citep[e.g.][]{Jimenez1997, BenitezLlambay2013, Benitez2017, Nebrin2023, Moreno2026}. 
Dark galaxies are predicted be found in both isolated environments \citep[e.g. REionization-Limited \HI\ Clouds (RELHICs),][]{Benitez2017} as well as around more massive hosts \citep[e.g. HI-rich Dark galaxiEs in Simulations (HIDES),][]{Zheng2025}. They should have relatively round and symmetric morphologies that are driven by the potential of their host dark matter, which may distinguish them from pressure-supported \HI\ clouds produced by ram pressure or tidal stripping events. 

Despite concerted efforts over the years \citep[e.g.][]{Minchin05,Haynes07} observational confirmation of dark galaxies remains challenging, with only a handful of promising candidates known to date \citep[e.g.,][]{Adams2013, Kwon2025, Li2025}. A starless \HI\ feature around M94 called Cloud-9, discovered by the Five-Hundred-Meter Aperture Spherical Telescope \citep[FAST;][]{Nan2011}, has been a particular focus of study in recent years \citep{Zhou2023, Benitez2023, Karunakaran2024, Benitez2024, Anand25, Zhou2026}. Deep optical observations place tight upper limits on presence of stars in Cloud 9 consistent with expectations for a dark galaxy, though its asymmetric morphology favors its interpretation as a pressure-supported cloud. High sensitivity \HI\ surveys are therefore needed to better sample the gas-rich dark galaxy population across varying environments in order to differentiate tidal dwarf galaxies from \HI\ reservoirs of cosmological origin and confirm the existence of dark galaxies \citep{Moreno2026}.

At the high-mass end of the HIMF, massive dark matter halos become exceedingly rare as their formation requires large overdense regions and long cosmic times \citep[e.g.,][]{PressSchechter1974}. Additionally, the gas reservoirs of most massive galaxies begin quenching by $z \sim 1$--$3$ through heating from the surrounding halo and feedback from active galactic nuclei (AGN) which contributes to the steep exponential cutoff of the HIMF observed in the local universe \citep[e.g.,][]{Croton2006, Dekel2006, Birnboim2007, Fabian2012, Bongiorno2016}. Gas-rich massive galaxies ($\MHI \gtrsim 10^{10} \solM$) at low redshifts are therefore uncommon and are interesting as many giants are found to have high gas fractions, hosting more \HI\ than would be expected from their stellar disks. For example, 34 sources cataloged in the `HIghMass' sample \citep{Huang2014} and 13 sources detected in the \HI\ eXtreme (HIX) survey \citep{Lutz2017} are found to have high gas fractions. This suggests unusual physical conditions allowing for more efficient gas accretion from their environment or less efficient star formation than typical galaxies \citep[e.g.,][]{Li2012}. While these gas-rich giants are \HI\ bright, a population census of these sources remains incomplete due to their sparsity beyond $\MHI > 10^{10.5} \solM$. Large surveys volumes ($z > 0.1$) are therefore needed to further constrain their number densities and properties. 

Previous untargeted \HI\ surveys such as the \HI\ Parkes All Sky Survey \citep[HIPASS;][]{Barnes2001, Meyer2004}, the Arecibo Legacy Fast ALFA \citep[ALFALFA;][]{Giovanelli2005, Haynes2018} and the currently ongoing FAST All Sky \HI\ survey \citep[FASHI;][]{Zhang2021, Zhang2024, FASHI_DR2} have made tremendous strides in constraining the gas content of luminous galaxies in the local universe ($z < 0.1$). Beyond the local universe, a few deep untargeted surveys such as the Blind Ultra-Deep HI Environmental Survey \citep[BUDHIES;][]{Verheijen2007}, the Arecibo Ultra-Deep Survey \citep[AUDS;][]{Hoppmann2015}, the COSMOS \HI\ Large Extragalactic Survey \citep[CHILES;][]{Hess2019} and the FAST Ultra-Deep Survey \citep[FUDS;][]{Xi24} have made sparse detections for $z>0.15$. Furthermore, the $0.1 < z < 0.3$ region is challenging for observing due to contamination from radio frequency interference (RFI) \citep[e.g.][]{MIGHTEE2021} and robust detections beyond $z>0.3$ remain limited. Only a handful ($\sim 10$) \HI\ observations at high redshifts ($z > 0.2$) have been reported so far \citep{Xi24,MIGHTEE_highz}. However, due to sensitivity and volume limits of these surveys, the HIMF has only robustly been constrained within $10^{7} \lesssim \MHI/\solM \lesssim 10^{10.5}$. A population census of gas-rich dwarfs and massive galaxies beyond these masses does not currently exist and the redshift evolution of the HIMF remains highly uncertain. 

To this end, the Canadian Hydrogen Observatory and Radio-Transient Detector (CHORD) is a next-generation radio telescope that is preparing to carry out one of the largest and deepest untargeted \HI\ survey to date \citep{CHORDWhitePaper}. Building on the expertise gained from the Canadian Hydrogen Intensity Mapping Experiment \citep[CHIME:][]{chimeOverview}, CHORD is currently under construction at the Dominion Radio Astrophysical Observatory (DRAO) and is designed to do precision 21 cm cosmology, discover fast radio transients and pulsars, and carry out widefield spectral line surveys \citep{CHORD2026}. The CHORD Pathfinder will consist of 64 dishes, followed by a full core array of 512 dishes. With 14,400 m$^{2}$ of instantaneous collecting area and a multi-year survey plan, CHORD has the potential to undertake the deepest widefield \HI\ survey in the Northern hemisphere in the upcoming years.

In this work, we discuss the \HI-galaxy science capabilities of CHORD, particularly in improving the population statistics of dwarf ($\MHI < 10^{7} \solM$) and massive ($\MHI > 10^{10.5}\solM$) galaxies to constrain the low-mass and high-mass end of the HIMF. Since CHORD is the first instrument to conduct an extragalactic survey with a maximally-redundant interferometer, we additionally address observational considerations specific to this design. The paper is structured as follows. In Section \ref{sec:CHORD}, we provide an overview of the CHORD instrument and observing strategy. In Section \ref{sec:Methods}, we outline our methodology for forecasting the CHORD survey outcomes with mock-\HI\ catalogs and spectra. The survey forecasts are subsequently presented in Section \ref{sec:Results}. The broader implications for \HI\ galaxy science are then contextualized in Section \ref{sec:Discussion}.

\section{\HI\ Galaxy Surveys with CHORD}
\label{sec:CHORD}

\begin{figure}
    \centering
    \includegraphics[width=\linewidth]{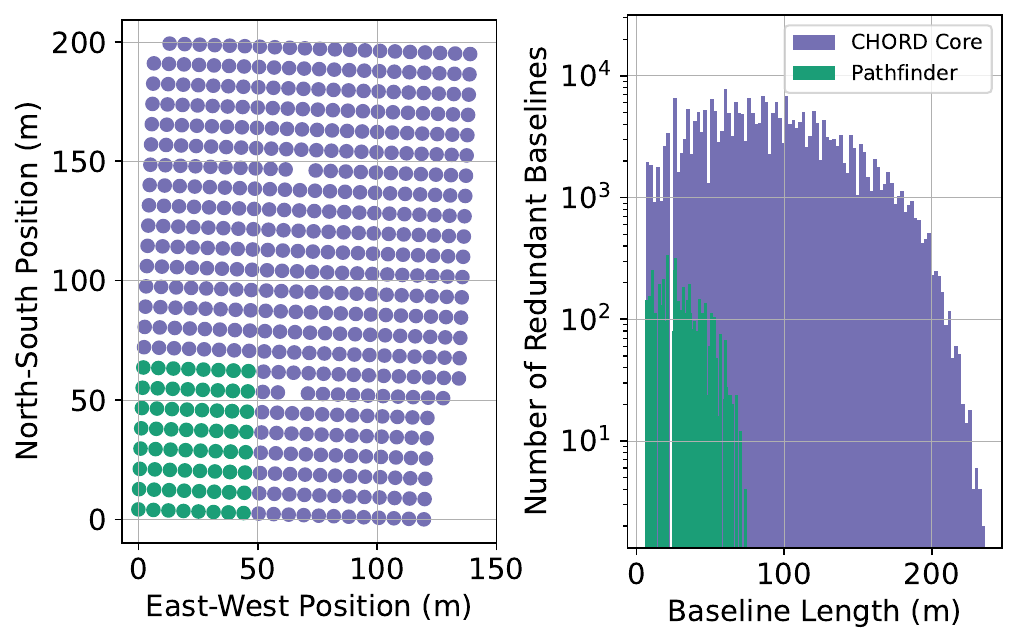}
    \caption{\textit{Left: } A schematic of the CHORD dish layout. Each 6-m wide dish is evenly spaced by 6.3\,m in the E-W direction and 8.5\,m in the N-S direction. The layout of the 64-dish initial Pathfinder is shown in green, while the full 512-dish core array (including the Pathfinder) is shown in purple. The dish positions follow the Universal Transverse Mercator (UTM) coordinate grid, which is rotated by 2$^{\circ}$ relative to geographical latitude and longitude at the CHORD location. The core array grid is not fully populated due to physical obstructions on the site.  
    \textit{Right: } Baseline distribution for the Pathfinder and core array, showing CHORD's high redundancy.}
    \label{fig:CHORD_layout}
\end{figure}

A layout of the CHORD core array is shown in the left panel of Figure \ref{fig:CHORD_layout}, where the dishes are arranged in a roughly regular rectangular grid. The right panel of Figure \ref{fig:CHORD_layout} shows the baseline distribution, highlighting the CHORD design choice to maximize redundancy. This is a significant departure from the irregular array layout of most interferometers that undertake \HI\ surveys \citep[e.g.\ ][]{HISurveys}, which are instead designed to maximize the number of unique baselines. While unique baselines have the advantage of probing different angular scales on the sky, redundant baselines improve sensitivity and dynamic range at the scales required for cosmology through intensity mapping \citep[e.g.,][]{Dillon2020}, which is a key CHORD science driver \citep{CHORDWhitePaper, CHORD2026}. 

Table \ref{tab:CHORD_params} summarizes the specifications of the CHORD core array at the \HI\ rest frequency of 1420\,MHz that we adopt in making our forecasts. For \HI\ science, the 300--1500\,MHz band corresponds to source redshifts in the range $0 \lesssim z \lesssim 3.7$, where the top of the band will be used for surveying \HI\ in individual galaxies, and the bottom of the band used for \HI\ intensity mapping. Each CHORD dish is $d=6\,$m in diameter, with $\eta_i\sim\,50\%$ illumination efficiency. We adopt the system temperature specification of $T_{\rm{sys}}=30\,$K, and the simulated performance of the CHORD ultra-wideband feeds return slightly lower values across the band of interest \citep{Mackay2023}. The full-width at half-maximum (FWHM) primary beam size at 1420 MHz is estimated as $\theta_{\rm{PB}} \approx 1.2\lambda/d = 2.4^{\circ}$, consistent with early data from the first CHORD dishes on site which measured the beam using transiting bright sources. The naturally-weighted synthesized beam FWHM at boresight is $\theta_{\rm{PB}}=4$\arcmin$\times6$\arcmin. The native spectral resolution of CHORD is 195.3 kHz, which corresponds to 42 \kms\ at $z=0$. However, spectra can be upchannelized to higher resolution using a two-stage polyphase filter bank (PFB). For the \HI\ galaxies science case, we henceforth assume a spectral resolution of $\Delta \nu_{\rm{chan}} = 23.7\,$kHz, which corresponds to 5 \kms\ at 1420 MHz.  
\begin{table}
    \centering
    \caption{CHORD Design Specifications at 1420 MHz}
    \begin{tabular}{cc}
        \hline 
        \vspace{1mm}
        &  \\
        \hline \hline
        Core dishes, $N_{\rm{dish}}$ & 512 (22$\times$24)  \\
        Dish diameter, $d$ & 6\,m  \\
        Grid spacing & 6.3\,m E-W, 8.5\,m N-S \\
        Bandpass & 300--1500 MHz \\
        Primary beam FWHM, $\theta_{\mathrm{PB}}$ & 2.4$^{\circ}$ \\
        Synthesized beam FWHM, $\theta_{\mathrm{SB}}$ & 4\arcmin\ N-S, 6\arcmin\ E-W \\
        Coarse spectral resolution & 195.3\,kHz / 42\, km s$^{-1}$ \\
        Fine spectral resolution, $\Delta \nu_{\rm{chan}}$ & 23.7\,kHz / 5\, km s$^{-1}$\\
        System temperature, $T_{\mathrm{sys}}$ & 30\,K \\
        Illumination efficiency, $\eta_{i}$  & 50\% \\
        Effective Dish Area, $A_e$ &  14\,m$^{2}$ \\
        Maximum survey footprint & 13200 deg$^{2}$ \\
        Accessible declinations & $+20^{\circ} < \delta < +80^{\circ}$ \\
         \hline
    \end{tabular}
    \label{tab:CHORD_params}
\end{table}

CHORD is a drift-scan telescope, where the dishes take data while parked at a fixed elevation and azimuth. As the Earth rotates, a strip of the celestial sphere is mapped continuously along the right ascension (RA) direction, such that the strip width in declination is roughly set by the primary beam FWHM. Integration time in the strip is accumulated by observing over multiple days. A full survey is then constructed by re-pointing the dishes to different elevations. Considering the $\sim5$\arcmin\ angular resolution of the CHORD core array synthesized beam, the vast majority of its individual \HI\ detections will be spatially unresolved. 

Given its redundant baseline design, source finding in CHORD survey data requires differentiating true emission-line sources from an ensemble of spatial aliases of that source since the sky is not Nyquist-sampled. In addition, the PFB upchannelization scheme produces spectral aliases that depend on the upchannelization factor. A visibility-based matched filtering search strategy for CHORD is presented in \cite{Hopkins2026}, who demonstrate that while aliases and true sources cannot be distinguished in an instantaneous snapshot, they can be disambiguated in the time-integrated case. Additionally, spacing the centers of individual declination strips by about a primary beam ($\sim2$--3$^{\circ}$) reduces the probability of mislocalizing a source, or mis-identifying an alias as a distinct source, to a negligible amount. Consistent with the results of \cite{Hopkins2026}, we forecast survey detections assuming that an implementation of this matched filter algorithm eliminates aliasing errors.  

The declination range accessible to CHORD is $+20^{\circ} < \delta < +80^{\circ}$, set by the range of dish elevations across which shadowing is insignificant for \HI-galaxy science. The total achievable survey footprint is therefore $\sim$13,000 deg$^{2}$ in the Northern sky. The CHORD operations plan includes a 5-year widefield survey \citep{CHORDWhitePaper} which this work assumes will cover the entire footprint. Given its significant collecting area, a 5-year survey with CHORD is nominally sensitive to individual \HI\ detections in the redshift range $0 \lesssim z \lesssim 1$. We therefore proceed to forecast survey detection statistics in this range for the fiducial CHORD 5-year survey period as well as for an initial 1-year CHORD campaign. 

Our scientific emphasis is on the census of the galaxy population that is achievable with these surveys, and therefore optimised to maximize sky area instead of depth across fewer declination strips \citep{HISurveys}. We restrict our forecasts to spatially unresolved detections. As we will demonstrate, practical considerations such as radio frequency interference (RFI) and source confusion hone these forecasts on the prospects for detecting a) low-mass nearby \HI\ sources, and b) high-mass distant ones. We describe our method for generating mock surveys in Section~\ref{sec:Methods}, and present our forecasts of these science cases in Section~\ref{sec:Results}.

\section{Methods}
\label{sec:Methods}

\begin{figure*}
    \centering
    \includegraphics[width=0.9\linewidth]{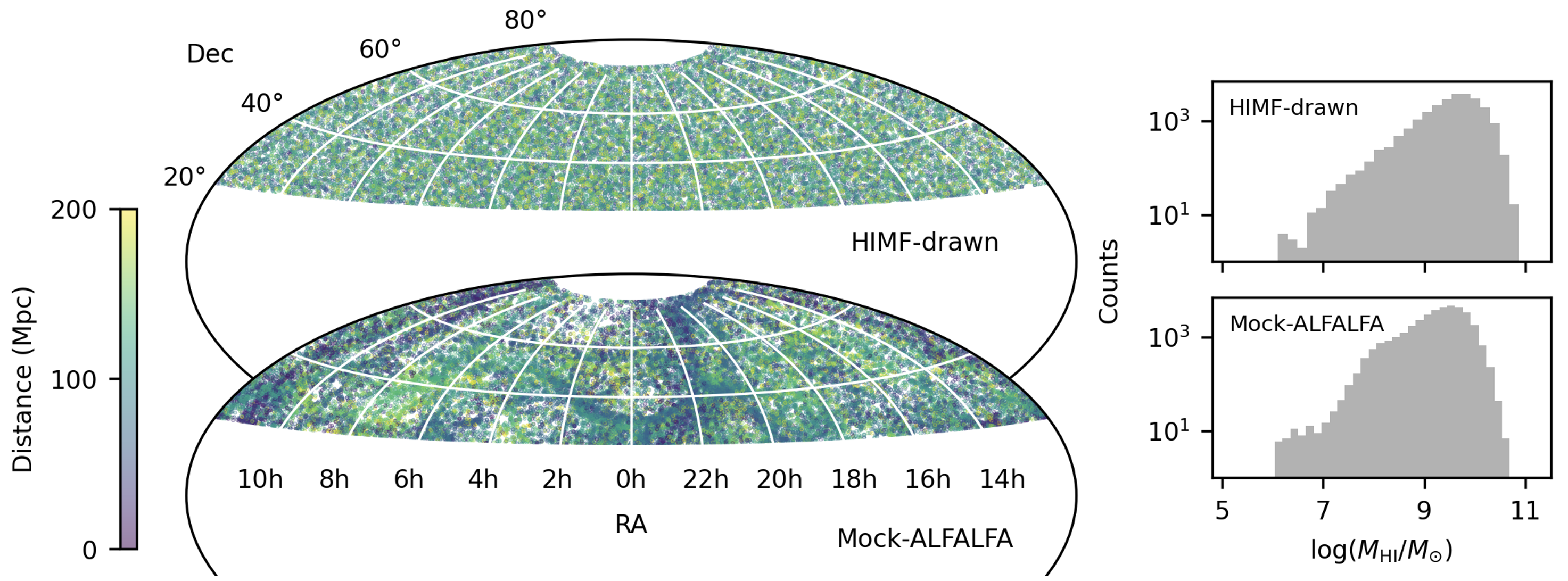}
    \caption{Comparison of mock \HI\ catalogs considered in this study.  The top map shows an HIMF-drawn mock-\HI\ catalog using the \citet{ALFALFA_Jones2018} parameters. We use HIMF-drawn catalogs to generate CHORD forecasts. The bottom map shows the mock-ALFALFA catalog from \cite{Brooks2023},  based on the \textsc{sibelius-dark} constrained simulation \citep{SibeliusDark}, that we use to assess the potential impact of cosmic variance on our forecasts (see Section \ref{sec:CosmicVariance}). The top map shows an ALFALFA-depth realization out to $200\,$Mpc (as opposed to a CHORD depth and volume mock sky) to enable direct comparison with the bottom panel. In both panels, each point represents a galaxy and the point color shows its distance. The right panels show the \MHI\ histogram of the sources included in the catalogs, highlighting differences between the catalogs at the low-mass end of the sample distribution at these survey depths.}
    \label{fig:catalogs}
\end{figure*}

In this section, we outline the methodology used for forecasting survey detections with CHORD: we describe our prescription for generating mock galaxy catalogs in Section~\ref{Sec:GenCat}, and spectral profiles of those galaxies in Section~\ref{sec:HISpectra}. We then describe how we construct samples of detections for fiducial 1-year and 5-year surveys with CHORD in Section~\ref{sec:constructSurvey}. 

\subsection{Mock \HI\ Catalogs}
\label{Sec:GenCat}
We forecast CHORD detections by constructing galaxy catalogs drawn from the ALFALFA HIMF measured by \cite{ALFALFA_Jones2018}:
\begin{equation}
\label{eq:Schechter}
    \phi(\MHI) = \frac{dn}{d\mathrm{log}(\MHI)} = \mathrm{ln}(10) \phi_{*} \left(\frac{\MHI}{\Mstar}\right)^{\alpha + 1}e^{\frac{\MHI}{\Mstar}},
\end{equation}
where the low-mass slope parameter is $\alpha=-1.25\pm0.02\pm0.1$, the characteristic mass or `knee' is log$(\Mstar h_{70}^{2}/\solM)=9.94\pm0.01\pm0.05$ and the normalization factor is $\phi_{*} = (4.5\pm0.2\pm0.8) \times 10^{-3} \, \Mpcdex$, where both the random and systematic uncertainties are given. For each draw from the HIMF, the fit parameters are sampled from a normal distribution within the total quoted errors. 

Throughout this work, we adopt a flat $\Lambda$CDM cosmology with \mbox{$\Hnot=70$ km s$^{-1}$ Mpc$^{-1}$} (i.e.\ $h_{70}=1$, consistent with \citealt{ALFALFA_Jones2018}) and $\Omegam= 0.315$.

Even though the ALFALFA HIMF was measured from detections in the mass range $ 7 \lesssim \log \MHI/\solM \lesssim 10.5$, we draw sources from the Schechter fit with masses as low as $M_{0} = 10^{5}\solM$ and as high as $M_{1} = 10^{12}\solM$ for the purposes of making forecasts. The implied number density of sources is therefore $n = \int_{M_0}^{M_1} \phi(\MHI) \, d\mathrm{log}(\MHI) \sim0.3\,\mathrm{Mpc}^{-3}$.  The catalog is generated by choosing a maximum redshift $z_{\rm{max}}$ for the survey volume $V_{\rm{max}}$, which is then split into co-moving $dV$ shells corresponding to equal redshift increments $dz$. In each shell, we draw an intrinsic \MHI\ population of $N= n dV$ samples. We adopt the same HIMF parameters at all source redshifts, preserving the number density in a unit co-moving volume. Sources drawn from the HIMF are assigned a random sky location, and their luminosity distance $D_L$ is computed from $z$ for the adopted cosmology. The galaxy inclination $i$ is sampled from a $\sin i$ probability distribution by drawing a uniformly distributed random value for $0 \leq \cos i \leq 1$, consistent with \cite{Brooks2023}.

To explore the effects of spatial clustering due to local large scale structure within the CHORD footprint, we also consider an ALFALFA-depth mock catalog constructed by \cite{Brooks2023} from the \textsc{sibelius-dark} constrained simulations \citep{SibeliusDark} that are discussed in more detail in Section \ref{sec:CosmicVariance}. Figure \ref{fig:catalogs} shows a realization of an ALFALFA-depth HIMF-drawn catalog (top map and top-right histogram) and the \textsc{sibelius-dark} ALFALFA-depth catalog that includes spatial clustering (bottom map and bottom-right histogram) to illustrate the large-scale structure within the CHORD footprint.

For a galaxy with an \HI\ mass \MHI, we select its rest-frame disk rotational velocity amplitude \VHI\ using the third-order polynomial fit to the $\MHI-\VHI$ relation determined by \citep{Lewis2019} and incorporated into the {\sc MCGSuite} package \citep[e.g.\ ][]{deg25}:
\begin{align}
\label{Eqn:MHI_VHI}
\log(\VHI/\kms) &=  0.0345 \xHI^{3} - 0.955 \xHI^2 \nonumber \\ 
&+ 9.13 \xHI - 28.0 \,\,\,,
\end{align}
where $\xHI \equiv \log(\MHI /\solM)$. The \MHI--\VHI\ relation was estimated by abundance matching \MHI\ samples drawn from the ALFALFA $\alpha.40$ HIMF derived in \cite{Martin10} and \VHI\ samples drawn from the (inclination-corrected) velocity function derived in \cite{Papastergis15}. Although we extrapolate this relation below the mass range $\MHI \gtrsim 10^{7} \solM$ used to calibrate it, this has little impact on our results because lower-mass systems are dispersion-dominated. We also find similar source count forecasts when drawing mock-catalogs from the extrapolated ALFALFA mass-width function \citep[MWF;][]{Papastergis15} consistent with \cite{Jones2015, Zhang2021, Oman2022}. 

The basic properties of galaxies in the mock-\HI\ catalogs are therefore \MHI, \VHI, $i$, $z$, $D_L$, RA and $\delta$, from which we estimate spectral profile shapes and detectability as described below.

\subsection{Mock HI Spectra}
\label{sec:HISpectra}

\begin{figure}
    \centering
    \includegraphics[width=\linewidth]{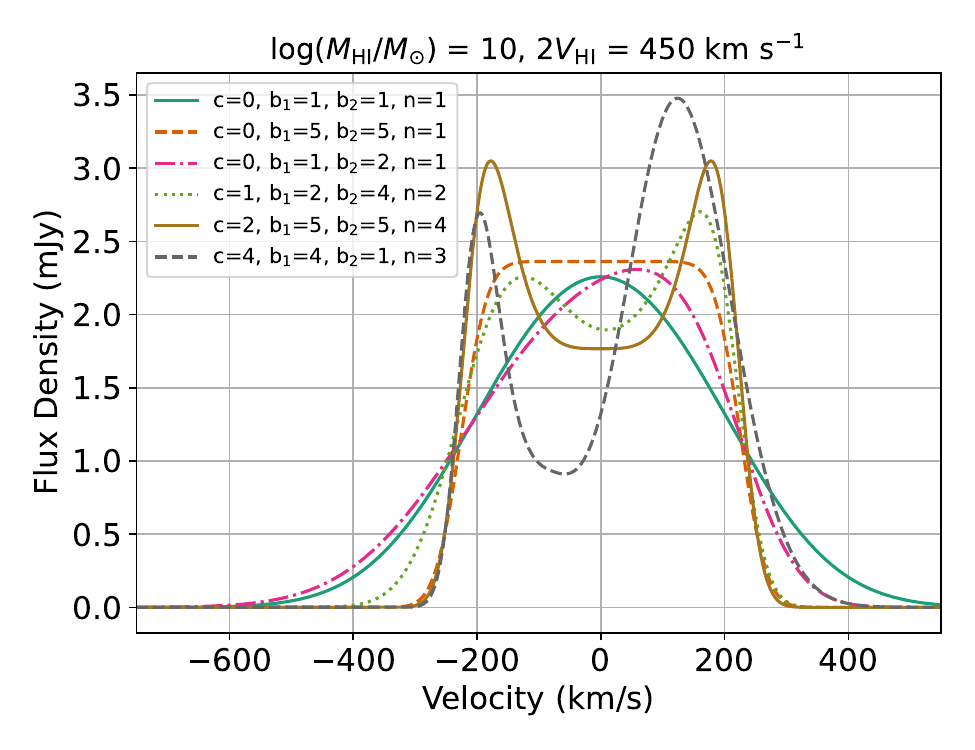}
    \caption{Diversity of \HI\ spectral profile shapes in the catalog, illustrated by different busy function representations of a single edge-on galaxy with a mass of $\MHI=10^{10} \solM$ and a rest frame spectral velocity width of $2\VHI=450\,\rm{km\,s}^{-1}$. 
    The spectra are produced by varying the busy function parameters $c, \, b_{1}, \, b_{2}$ and $n$ within the adopted ranges. The busy function parameters related to the spectral height and width are set to $a=w=1$, and are subsequently scaled to physical units as described in the text.}
    \label{fig:Busy_shapes}
\end{figure}

For each galaxy in the catalog, we generate a mock \HI\ spectral profile with a randomized shape using the analytical generalized busy function from \cite{BusyFunction}:
\begin{multline}
\label{Eqn:Busy}
B_{s}(x) = \frac{a}{4} \times 
(\mathrm{erf}[b_{1}\{w + x\}] + 1) \times \\
(\mathrm{erf}[b_{2}\{w - x\}] + 1) \times
(cx^{n} + 1), 
\end{multline}
where $a, \, b_1, \, b_2, \, c, \, w, \, n$, are free parameters and $x$ represents the spectral axis for the data. In Equation \ref{Eqn:Busy}, the polynomial term shapes the central trough of the profile, while the error function terms set the profile flanks. We set the properties of the central profile trough by uniformly sampling the trough depth parameter $c$ in the range 0--4 and the trough shape parameter $n$ in the range 1--4. We then select the parameters $b_1$ and $b_2$ which modify the slope of the low velocity and high velocity flanks, respectively, by uniformly sampling values between 1--5. Figure \ref{fig:Busy_shapes} demonstrates the diversity in spectral profiles that can be generated for a single galaxy for our adopted ranges of $c, \, n, \, b_1, \, b_2$: as is the case for known \HI-galaxy detections, the profiles can be Gaussian-like, boxy, double-horned, symmetric or asymmetric. 

With the spectral shape determined, we then set the width parameter $w$ to produce a busy function FWHM that corresponds to the rotationally-broadened component of the rest-frame profile, $2\sin i \, \VHI$, where \VHI\ is determined from \MHI\ from Equation~\ref{Eqn:MHI_VHI}. The height parameter $a$ is set so that the profile integrates to a rest-frame flux integral \mbox{$S^{V_{\mathrm{rest}}}=\int S\,dV$} that is consistent with that source's \MHI\ and $D_L$  \citep[e.g.\ ][]{Meyer2017}:
\begin{equation}
    \label{Eqn:MHI_S21}
    \frac{\MHI}{\solM} \simeq \frac{2.36 \times 10^5}{(1+z)} \left(\frac{D_L}{\mathrm{Mpc}}\right)^2 \left(\frac{S^{V_{\mathrm{rest}}}}{\mathrm{Jy \, \kms}}\right). 
\end{equation}
To model the effects of thermal and turbulent broadening, we convolve each rest-frame mock spectrum with a normalized Gaussian kernel with a standard deviation of $10\ \kms$, consistent with the average measured velocity dispersion in the outskirts of real \HI\ disks  \citep[e.g.\ ][]{Tamburro2009}. This results in a rest frame spectral width of
\begin{equation}
\label{Eqn:W50_broad}
    \Wfifty \sim \sqrt{(2\sin i\,\VHI)^{2} + (2\sqrt{2\ln2}\,10)^{2}}.
\end{equation}

\begin{figure*}
    \centering
    \includegraphics[width=0.9\linewidth]{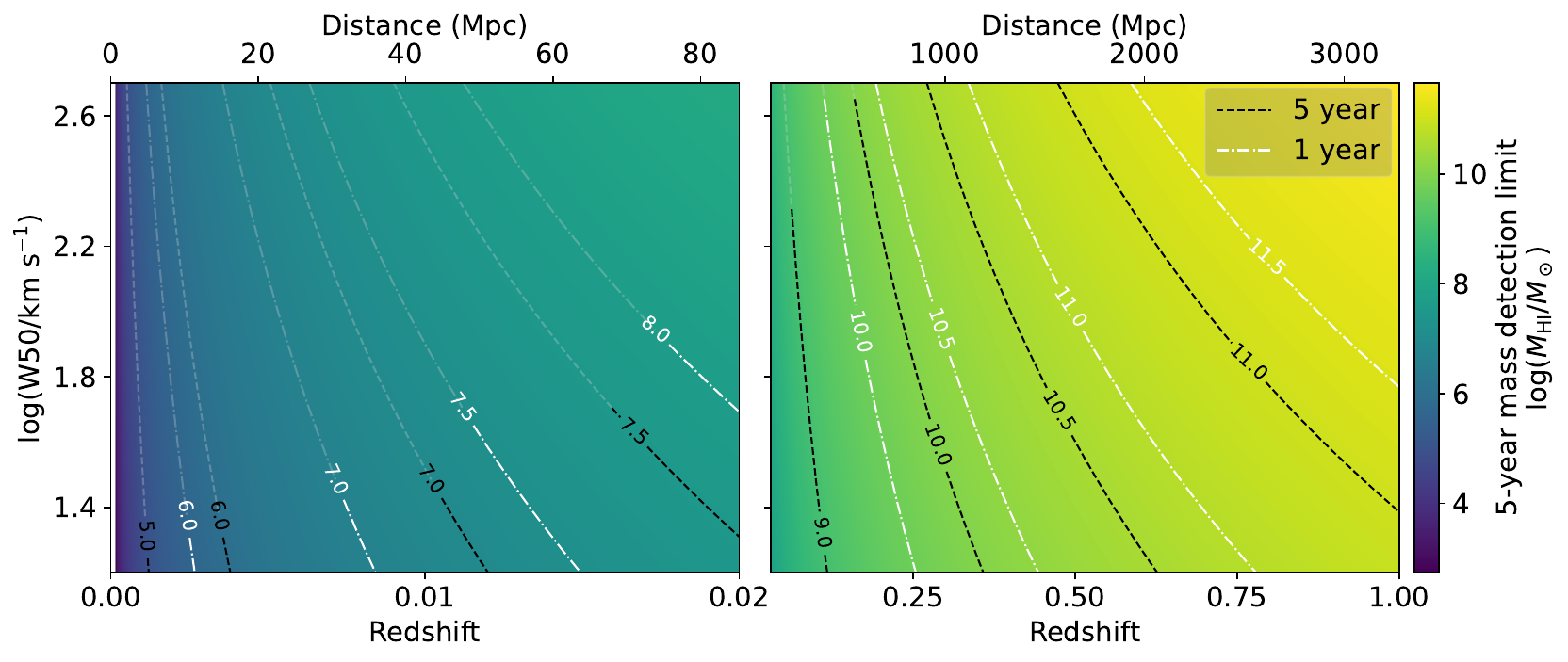}
    \caption{The minimum $\MHI$ detectable with CHORD at a given distance and observed rest frame spectral width $W_{50}$ using an integrated $S/N_{\rm{int}}=6$. The panels highlight the low-redshift (left) and high-redshift (right) extremes of the detected samples. The colormap is consistent between the panels, and shows the logarithm of $\MHI$ detectable for a 5-year survey. The white and black contours are labeled by logarithmic mass for a 1-year survey and 5-year CHORD survey, respectively, and are drawn up to the maximum expected \Wfifty\ for an edge-on source with \VHI\ from equation~\ref{Eqn:MHI_VHI}. The gray contours that follow for higher \Wfifty\ convey that sources are not expected to occupy this parameter space. Sources with a given $\MHI$ and $W_{50}$ along a contour are detectable at lower distances/redshifts and/or lower $W_{50}$.}
    \label{fig:Dectability}
\end{figure*}

For dispersion-supported, low-mass and/or nearly face-on galaxies where the rotational broadening term is small, this imposes a minimum spectral FWHM of $\sim 25$ \kms, consistent with observations. The convolution also smooths over any narrow features in the busy function shape, resulting in the characteristic Gaussian-like spectra expected for dwarfs and face-on systems. The additional width broadening from (redshift-dependent) instrumental effects is relatively small, and we therefore omit them from our forecasts for simplicity. 

To determine the detectability of a given mock spectrum, we estimate the noise per spectral channel of width $\Delta\nu_{\mathrm{chan}}$, $\sigma_{\rm{}chan}$, from the radiometer equation: 
\begin{equation}
\label{Eqn:radiometer}
    \sigma_{\rm{chan}} = \frac{ 2k_{B}T_{\mathrm{sys}}}{A_{e}\sqrt{N_{\mathrm{dish}}(N_\mathrm{dish}-1)\Delta\nu_{\mathrm{chan}}\tau_{\mathrm{eff}}}},
\end{equation}
where dual polarization and natural weighting are assumed. Here, $k_B$ is the Boltzmann constant and $\tau_{\mathrm{eff}}$ is the effective integration time that is described in the next section. The CHORD specifications for the system temperature $T_{\mathrm{sys}}$, effective collecting area $A_e = \eta_i\pi (d/2)^2$, and number of dishes $N_{\mathrm{dish}}$ are given in Table \ref{tab:CHORD_params}. 

Next, we calculate the integrated signal-to-noise $S/N_{\rm{int}}$ for each spectrum by integrating the spectral flux density $S_{\nu}$ along the frequency axis over $N_{\rm{chan}}$ channels as in \cite{Meyer2017}:
\begin{equation}
    \label{Eqn:SNR_int}
    S/N_{\rm{int}} = \frac{\int S_{\nu}\,d\nu}{\sigma_{\rm{}chan}\Delta\nu_{\rm{chan}}{\sqrt{N_{\rm{chan}}}}}.
\end{equation}
 
As confirmed by reliability tests of the detection schemes for other untargetted \HI\ surveys \citep[e.g.\ ][]{Saintonge2007, Wallaby2022}, we adopt $S/N_{\rm{int}}=6$ as our detection threshold for the purposes of generating \HI\ forecasts.   

\subsection{Constructing a Survey}
\label{sec:constructSurvey}

The drift-scan observing strategy of CHORD implies that the total time on source is aggregated by observing the same strip of sky over multiple days. To relate the effective integration time $\tau_{\mathrm{eff}}$ in equation \ref{Eqn:radiometer} to the total observation length in days, we estimate the time during which a source falls within the CHORD primary beam each day as:
\begin{equation}
\label{Eqn:tau_day}
     \tau_{d} = \frac{\theta_{\mathrm{PB}}(1+z)}{\omega\,\cos(\delta_0)},
\end{equation}
where $\omega$ is the Earth's sidereal rotation rate, $\delta_0$ is the pointing declination, and the $(1+z)$ factor evolves the primary beam FWHM $\theta_{\mathrm{PB}}$ as a function of redshift. For reference at $z=0$, a source at declination $\delta = 20^{\circ}$ will spend $\sim$10 minutes in the primary beam, while one at $\delta = 80^{\circ}$ will spend $\sim$55 minutes per day. Equation \ref{Eqn:tau_day} assumes there is no attenuation by the primary beam in the right ascension direction. Including this effect, if the data are imaged to the half power level and mosaicked using variance weights then the noise per channel would increase by about a factor of $\sim 1.2$.

The effective integration time is thus $\tau_{\mathrm{eff}}=\tau_{d}\cdot n_{d}$, where $n_d$ is the observation length in days. Combining equations \ref{Eqn:radiometer} and \ref{Eqn:tau_day} gives:
\begin{equation}
\label{Eqn:radiometer_drift}
    \sigma_{\rm{chan}} = \frac{ 2k_{B}T_{\mathrm{sys}}}{A_{e}}\sqrt{\frac{\omega\,\mathrm{cos}(\delta_0)}{{N_{\mathrm{dish}}(N_\mathrm{dish}-1)\Delta\nu_{\mathrm{chan}}n_d\,\theta_{PB}(1+z)}}}.
\end{equation}

For simplicity in this work, we forecast source counts assuming a survey of roughly uniform sensitivity across the survey footprint. This requires that the number of observing days $n_{d}$ scales with $\cos(\delta_{0})$ such that more time is spent at lower declinations compared to higher declinations to achieve equal depth. We also account for the time required to repoint the dishes to different elevations to cover the footprint, which we assume to be one week. We allocate the remaining survey time to observations at each pointing, scaled by $\cos\delta_{0}$ to achieve roughly uniform depth. For a 5-year survey, we assume that 25 pointings spaced by $2.5^{\circ}$ are used to cover the full $+20^{\circ} \leq \delta \leq +80^{\circ}$ footprint accessible to CHORD. We ignore the attenuation by the primary beam in the declination direction for simplicity, which in practice will depend on the exact spacing between pointings, and in this case results in a $\sim5\%$ variation in the beam response. 

Given the time required to repoint the dishes, we assume that a 1-year survey will cover a more limited sky area and adopt 13 pointings spaced by $2.5^{\circ}$ to cover the bottom half of the sky ($+20^{\circ} < \delta < +50^{\circ}$). With these parameters, we find that the sensitivity that can be achieved for a 1-year survey is $\sim$0.6 mJy per fine spectral channel with $\Delta \nu_{\rm{chan}} = 23.7\,$kHz, and a sensitivity of $\sim$0.3 mJy/channel for a 5-year survey. 

Based on these sensitivity estimates, we show the approximate \HI\ mass detection limit for CHORD surveys in Figure~\ref{fig:Dectability}. The contours show the minimum \MHI\ mass that can be detected for a given distance and spectral velocity width at rest, estimated for $S/N_{\rm{int}}=6$ by combining equations~\ref{Eqn:MHI_S21} and \ref{Eqn:SNR_int}:
\begin{equation}
\label{Eqn:massLimit}
   \MHI \simeq 2.36\times 10^{5}\,D_{L}^{2}\,\sigma_{\rm{chan}}\,S/N_{\rm{int}}\,\sqrt{\frac{c\,\Delta\nu_{\rm{chan}}\,\Delta V_{\rm{rest}}}{\nu_{\,\HI}(1+z)}}.
\end{equation}

\noindent Here $c$ is the speed of light, $\nu_{\,\HI}=1420\,$MHz is the \HI\ line rest frequency, and $\Delta V_{\rm{rest}}$ is the rest frame velocity width of the spectrum which we approximate as \Wfifty\ from equation~\ref{Eqn:W50_broad}. 
Figure~\ref{fig:Dectability} illustrates the well-known inclination dependence on the detectability of a source, with lower-inclination systems producing narrower $W_{50}$ that can be detected out to larger distances. Also as expected, galaxies of a given \HI\ mass are detectable to larger distances/higher redshifts in the 5-year survey than those in the 1-year survey. 
Figure~\ref{fig:Dectability}'s division by distance into local galaxies (left panel) and high-redshift sources (right panel) highlights the populations that CHORD will excel at detecting: it is sensitive to galaxies as small as $\log \MHI/\solM \sim 5$ in the very local volume, and giant gas-rich systems with $\log \MHI/\solM \sim 11$ out to $z \sim 1$. In the next section, we forecast how well CHORD will be able to constrain the low-mass and high-mass ends of the HIMF based on these sensitivity estimates. 

\section{Forecasting Results}
\label{sec:Results}
In this section, we present source count forecasts for a 1-year and 5-year survey with CHORD. We demonstrate that the expected detections will place new constraints on the low-mass end of the HIMF in Section~\ref{sec:HIMF_Constrain} and on the high-mass end of the HIMF in Section~\ref{sec:HighzForecasts}. The impact of spectroscopic confusion on survey outcomes is presented in Section~\ref{sec:Confusion}.

\subsection{Constraining the faint end of the HIMF}
\label{sec:HIMF_Constrain}

\begin{figure}[tbh!]
    \centering
    \includegraphics[width=\linewidth]{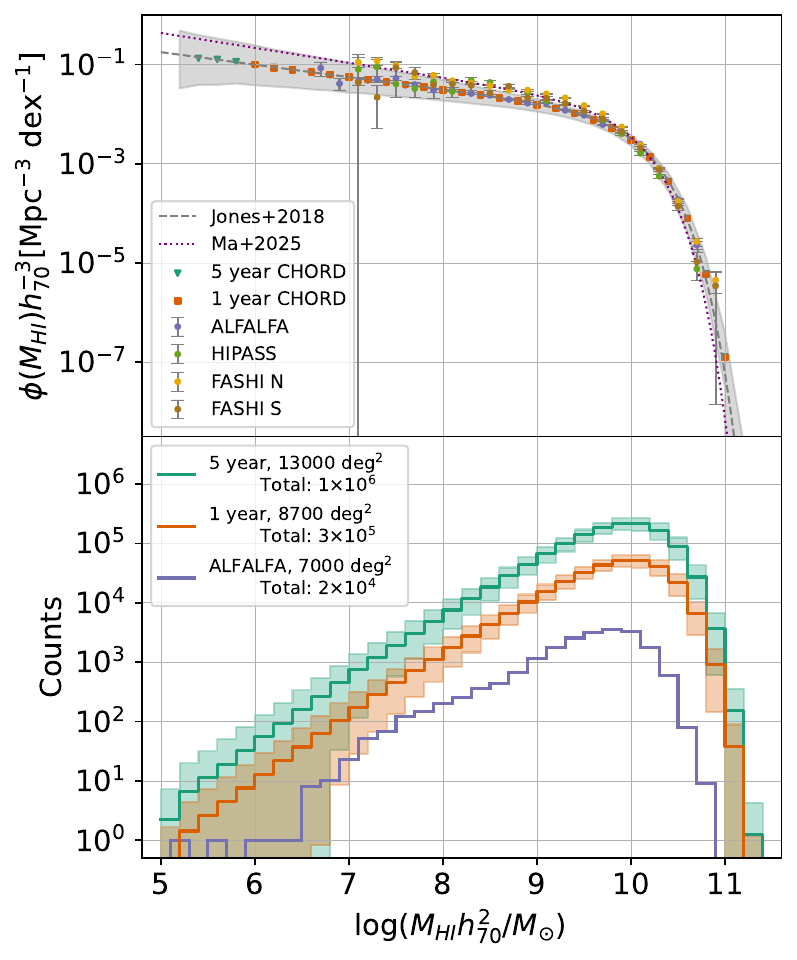}
    \caption{1-year and 5-year CHORD survey forecasts, ignoring RFI and source confusion. \textit{Top:} Median reconstructed HIMF, where only 0.2 dex mass bins with 10 or more detections are included. The gray shaded region shows the reconstructed HIMF variation between the $16^{\rm{th}}$ and $84^{\rm{th}}$ percentiles from 1000 catalog realizations. For comparison, the ALFALFA HIMF measurements from \cite{ALFALFA_Jones2018} are shown with the accompanying Schechter fit plotted in the dashed line. The HIMF measurements from HIPASS and FASHI DR1 are taken from Table 3 of \cite{Ma2025} and their Schechter fit is plotted in the dotted line. \textit{Bottom:} Forecast detection counts. The solid lines show the mean number of detections calculated from 1000 realizations of the HIMF-drawn catalogs, and the shaded regions show the standard deviation in each bin across the realizations. The purple line shows the ALFALFA $\alpha$.100 code 1 source counts, estimated to be $> 90\%$ complete and with heliocentric velocities of \mbox{$0 \, \mathrm{\kms} < cz < 15000 \, \mathrm{\kms}$} \citep{Haynes2018}.}
    \label{fig:HIMF_constrain}
\end{figure}

To forecast survey outcomes, we draw a thousand realizations of a mock-\HI\ sky within a $z\leq 1$ volume from HIMFs varied within the quoted errors of the \cite{ALFALFA_Jones2018} Schechter fit (described in Section~\ref{Sec:GenCat}) with randomized spectral profile shapes which are assessed for their detectability (described in Section~\ref{sec:HISpectra}) based on survey parameters (described in Section~\ref{sec:constructSurvey}). From the predicted detections for each catalog realization, we reconstruct the HIMF using the $\Sigma1/V_{\rm{max}}$ method, which weights each source count by the reciprocal of the maximum comoving volume $1/V_{\rm{max}}$ over which the source can be detected \citep[e.g.,][]{Schmidt1968, Martin10}. The weighted counts are then summed in each mass bin to estimate $\phi(\MHI)$, where only bins with 10 or more detections are considered. In practice, a more sophisticated method will be applied to real data to account for various observational uncertainties arising from measurement biases, cosmic variance, and survey incompleteness (e.g.\ the $1/V_{\rm{eff}}$ or generalized maximum likelihood estimators of \citealt{Zwaan2005} and \citealt{Obreschkow2018}, or the recovery matrix method from from \citet{Kazemi-Moridani2025}), but $1/V_{\rm{max}}$ is suitable for our HIMF-drawn forecasting approach. 

Figure~\ref{fig:HIMF_constrain} shows the detection forecasts and the reconstructed HIMF values (top) and source counts (bottom) for all catalog realizations. These initial forecasts ignore the effects of RFI contamination (addressed in Section \ref{sec:HighzForecasts}) and source confusion (addressed in Section \ref{sec:Confusion}). The median recovered HIMF lies directly on the \cite{ALFALFA_Jones2018} Schechter function, with the $16^{\rm{th}}$ and $84^{\rm{th}}$ percentile values following their quoted errors, as expected, since we use their fit parameters. We also include the distribution of ALFALFA $\alpha.100$ code 1 source counts from the 90\% completeness limit with heliocentric velocities of \mbox{$0 \, \mathrm{\kms} < cz < 15000 \, \mathrm{\kms}$} \citep{Haynes2018}, as well as the HIMFs reported from HIPASS \citep{Zwaan2005}, ALFALFA \citep{ALFALFA_Jones2018}, and the first release of the FASHI \citep{Ma2025} survey. 


Figure~\ref{fig:HIMF_constrain} illustrates that the 1-year and 5-year CHORD surveys are expected to observe one and two orders of magnitude more galaxies than ALFALFA $\alpha.100$, respectively, though the statistics in the mass range \mbox{$8.5 \lesssim \MHI/\solM \lesssim 10.5$} are impacted by RFI and confusion, as we explore below. The detection statistics imply that CHORD will expand our census of \HI-rich galaxies to significantly lower masses than previously probed. 

We forecast that even a 1-year survey will reach sufficient sensitivity (0.6 mJy at 5 \kms\ resolution) to push down the low-mass end of HIMF to $\sim 10^{6} \solM$. This is about an order of magnitude lower than the HIMF measurements from the surveys shown in the top panel of Figure~\ref{fig:HIMF_constrain} and comparable with the recently-released FASHI DR2 results \citep{FASHI_DR2} which robustly constrain the HIMF to $\sim 10^{6.2} \solM$. With a 0.3 mJy sensitivity, a 5-year survey will constrain the HIMF further towards $\sim 10^{5.5}\solM$, finding thousands of $10^6 < \MHI/\solM < 10^7$ sources and roughly $\sim 70$ Leo T type sources with \mbox{$10^5 < \MHI/\solM < 10^6$}. Approximately half of these Leo T mass detections are expected to be located closer than 5 Mpc, where line-of-sight velocities can overlap with High Velocity Clouds (HVCs) of the Milky Way and M31, and follow-up observations may therefore be needed to confirm extragalactic origin. Detection counts below $\MHI < 10^{7} \solM$ assume that the intrinsic number density distribution of galaxies still follows the Schechter relation. If instead there is a turnover in the low-mass end of the HIMF \citep[as predicted by][]{Kim2015, Baugh2019, Dave2020}, then CHORD can provide constraints of the mass at which this turnover occurs. This mass regime will directly probe the population of optically dim and dark galaxies testing theoretical predictions at the very edge of galaxy formation, as discussed in Section \ref{sec:DarkGalaxies}.

\subsection{High Redshift Forecasts}
\label{sec:HighzForecasts}

\begin{figure}[tbh!]
    \centering
    \includegraphics[width=\linewidth]{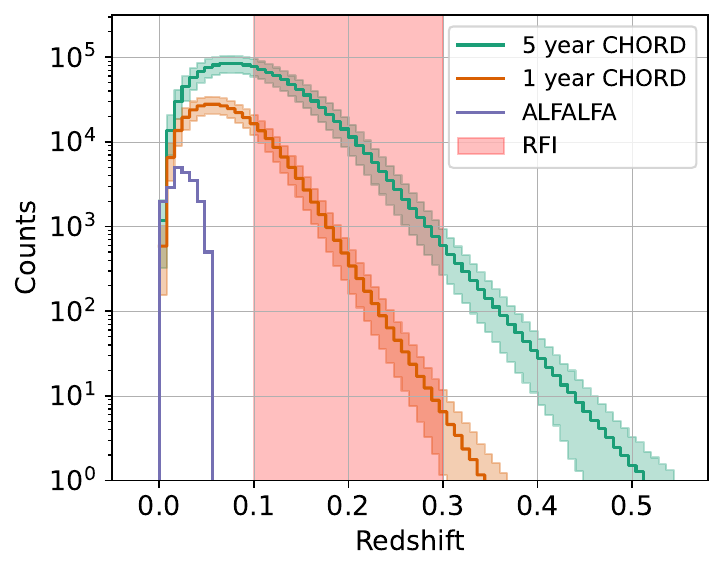}
    \caption{Predicted CHORD survey source counts as a function of redshift. The solid lines show the mean counts in $\delta z = 0.008$ bins across 1000 HIMF-drawn catalog realizations out to $z=1$. The shaded regions show the standard deviation in counts resulting from the HIMF uncertainties. The same ALFALFA $\alpha.100$ sources shown in Figure~\ref{fig:HIMF_constrain} are included here as a reference. The RFI-dominated band from 1090--1290 MHz, corresponding to a redshift range \mbox{$0.1 \lesssim z \lesssim 0.3$}, is shaded in red. These counts ignore source blending, which is discussed in Section \ref{sec:Confusion}.}
    \label{fig:Highz}
\end{figure}

Figure~\ref{fig:Dectability} implies that 1-year and 5-year CHORD surveys are sensitive to gas-rich giants $\MHI \gtrsim 10^{10.5} \solM$, and Figure \ref{fig:HIMF_constrain} suggests that CHORD will detect thousands of these systems. Although they are bright, the current census of high-\HI\ mass systems is limited by a combination of their rarity, the impact of RFI and the instrumental bandpasses of previous surveys. 
This is illustrated in Figure~\ref{fig:Highz}, which shows the detection counts as a function of redshift that we forecast for CHORD, in comparison to ALFALFA, with consideration of the range over which RFI is expected to be significant \citep[e.g.,][]{Offringa2023}.

Figure~\ref{fig:Highz} illustrates the significantly larger volume that CHORD will probe in relation to ALFALFA, for which the sharp drop in source counts beyond $z \sim 0.06$ stems from the ALFA bandpass \citep{Haynes2018}. CHORD's ultra-wideband feed enables searches out to a significantly larger volumes, increasing samples of giant \HI\ galaxies by more than an order of magnitude beyond current observations. These results are consistent with initial CHORD forecasts (see figure 3 of \citealt{CHORDWhitePaper}). Discounting the redshift range likely to be contaminated by RFI and considering the \HI\ mass detectability plot in Figure~\ref{fig:Dectability}, the majority of these massive galaxy detections will lie beyond $z \gtrsim 0.3$. Since uncertainties in the high-mass source counts from the HIMF are  higher than Poisson counting error (see \citealt{ALFALFA_Jones2018} for a full discussion on the sources of uncertainty), CHORD is expected to constrain the HIMF out to approximately $10^{11} \solM$. Furthermore, deviations from the local $z=0$ HIMF at high masses may provide preliminary indications of redshift evolution, which is currently uncertain \citep[e.g.,][]{Power2010, Bera2022,Arlow2026}. 

At this stage, we do not remove source counts between \mbox{$0.1 \lesssim z \lesssim 0.3$} from the total forecast estimates since there may be some potential to excise RFI in this band. Previous surveys relied on baseband dumps at a cadence of several seconds \citep[e.g.\ ][]{Haynes2018, Koribalski2020, MIGHTEE2021}, whereas CHORD's millisecond raw-voltage dumps and daily observations of the same sky may enable novel RFI excision approaches \citep[e.g.,][]{Mirhosseini2020PhDT}. Ongoing work is examining whether the GNSS duty cycle allows for more effective excision with the higher time-resolution data and dedicated RFI monitoring antennas on site.

\subsection{\HI\ Source Confusion}
\label{sec:Confusion}
Considering the relatively large size of the CHORD synthesized beam, it is important to consider the extent to which the forecasts in Sections~\ref{sec:HIMF_Constrain}~and~\ref{sec:HighzForecasts} are impacted by detections that are blends of more than one source. These blended detections can result in incorrect fluxes, masses, velocity widths and number densities. This can lead to biases in measuring the HIMF, which is central to CHORD \HI\ galaxy science. We therefore forecast the degree of spectroscopic confusion, which occurs when two or more sources simultaneously overlap spectrally and spatially within a beam width. 

To predict the rate of confusion for a 5-year CHORD survey, we use the spectroscopic confusion model from \cite{Jones2015}. In this model, the average number of additional sources in a beam-sized cylindrical volume around a given source is determined from the combined distributions of the HIMF and mass-conditional velocity width function, as well as a two-dimensional correlation function estimating the separation between sources (see \citealt{Jones2015} for a full derivation). A Monte Carlo integration over 1000 trials is used to numerically estimate the rate of confusion. The model assumes no redshift evolution of the HIMF, the \HI\ density of the universe and the \HI\ correlation function in comoving coordinates as these quantities are poorly constrained and some studies suggest a weak evolution in \HI\ \citep[e.g.,][]{Power2010, Arlow2026, Zhang2026}.  

Two sources are defined as confused if they subtend an angle smaller than the synthesized beam and the sum of their velocity widths is more than twice their redshift separation. The first condition is given by:
\begin{equation}
    \kappa < \theta_{SB}(1+z) \, D_{A},
\end{equation}
where $\kappa$ is the projected linear distance, $\theta_{SB}$ is the angular diameter of the synthesized beam at $z=0$, and $D_{A}$ is the angular diameter distance. The second condition is given by:
\begin{equation}
    \beta < \frac{W_{0}(1+z)}{2H_{0}},
\end{equation}
where $\beta$ is the line-of-sight separation between the sources, $W_0$ is the sum of their spectral velocity widths at rest and $H_{0}$ is the Hubble constant. $H_{0}$ is kept constant over the redshift interval as it used in the integration limit of the $z\sim0$ correlation function.

\begin{figure*}[tbh!]
    \centering
    \includegraphics[width=\linewidth]{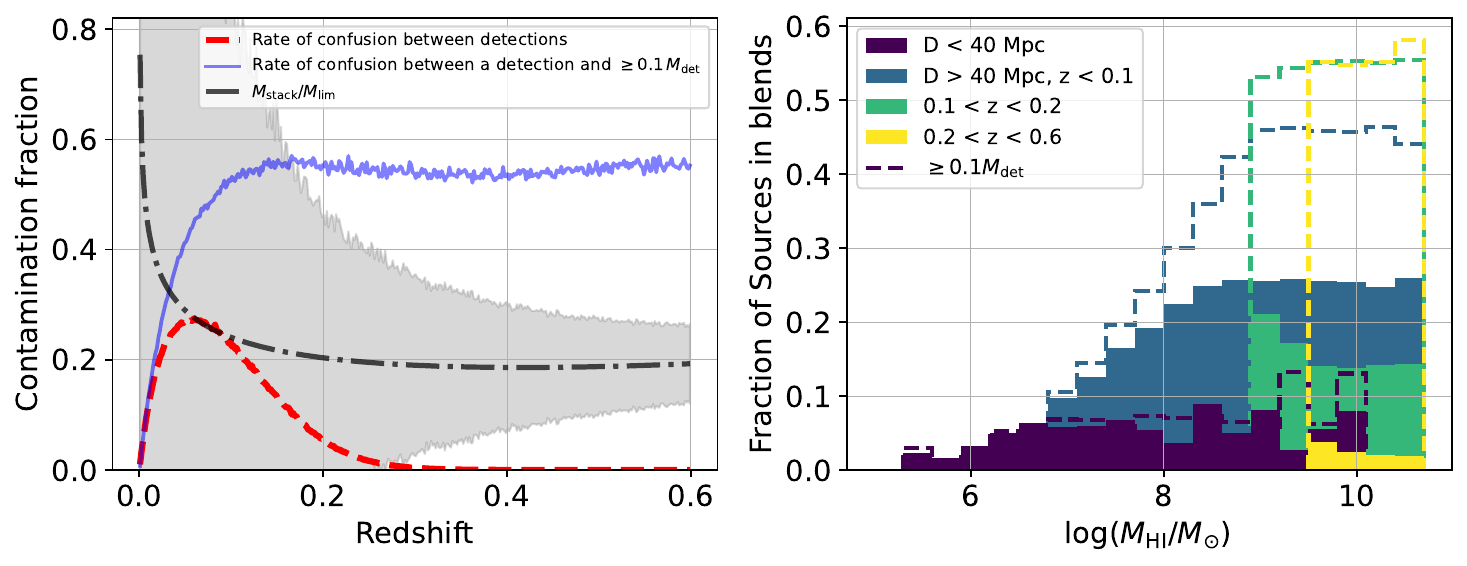}
    \caption{Anticipated impact of spectroscopic confusion in a CHORD 5-year survey. \textit{Left:} Confusion as a function of redshift, estimated from a Monte Carlo integration of 1000 sky realizations from the mass-width and correlation functions described in \cite{Jones2015}. The confusion rate between two or more individually detectable sources is shown by the red dashed line. The confusion rate between a detectable source with \HI\ mass $M_{\rm{det}}$ and at least one spectroscopically confused source with $\MHI > 0.1\,M_{\rm{det}}$ is shown by the blue solid line. The mean integrated \HI-mass of all confused sources ($M_{\rm{stack}}$) as a fraction of the \HI-mass detection limit ($M_{\rm{lim}}$) is shown by the black dot-dashed line. The gray shaded region shows the standard deviation in $M_{\rm{stack}}/M_{\rm{lim}}$ across the 1000 realizations. \textit{Right:} Confusion as a function of \HI\ mass for a single survey realization, with colors indicating different distance/redshift ranges. Confusion between two or more individually detectable sources in different distance and redshift ranges is shown by the solid histograms, while the confusion rate between a detectable source with \HI\ mass $M_{\rm{det}}$ and at least one spectroscopically confused source with $\MHI > 0.1\,M_{\rm{det}}$ is shown by the dashed histograms.}
    \label{fig:Confusion}
\end{figure*}

Following \citet{Jones2015}, we estimate two types of spectroscopic confusion rates. The first is the rate at which two or more individually detectable sources are spectroscopically confused; they register as a single brighter detection instead of individual detections because of their relative proximity. To account for confusion with sources below the sensitivity limit, we additionally estimate a second, more pessimistic rate at which an individually detectable source with \HI\ mass $M_{\rm{det}}$ is spectroscopically confused with at least one other source with that has $\MHI > 0.1\,M_{\rm{det}}$. Analogous to \citet{Jones2016}, we quantify the relative contamination by estimating the mean integrated \HI-mass ($M_{\rm{stack}}$) of all confused sources in the synthesized beam as a fraction of the \HI-mass detection limit ($M_{\rm{lim}}$). $M_{\rm{lim}}$ is estimated from Equation~\ref{Eqn:massLimit} for a nearly face-on system with the minimum at-rest spectral width of $\Delta V_{\rm{rest}} = 25$ \kms. Figure \ref{fig:Confusion} shows these confusion and contamination rate estimates for a 5-year CHORD survey as a function of redshift (left panel) and \HI\ mass (right panel). 

When considering confusion between two or more individually detectable sources, we find that the nearest ($D < 40$ Mpc) and furthest ($0.2 < z < 0.6$) survey volumes have confusion rates $< 5-10\%$. In contrast, $> 25\%$ of detections are expected to result from spectroscopically confused sources at intermediate distances ($D > 40, z < 0.2$). 

The low rate of confusion in the nearby universe is due both from the narrow line-widths of the predominantly low-mass sources being detected and the relatively small physical size of the synthesized beam. The physical beam grows linearly in diameter as the distance increases, and at intermediate distances only the more massive sources ($\MHI \gtrsim 10^{7} \solM$) with wider spectra are detectable, increasing the chances of both spatial and spectral confusion. At high redshifts ($z > 0.2$), even though the physical size of the synthesized beam is on the order of galaxy clusters ($\theta_{\rm{SB}}(1+z)\gtrsim 1$ Mpc), the chance of confusion between individually detectable sources is negligible since only the rarer massive sources beyond the knee of HIMF will lie above the sensitivity limit (see Figure~\ref{fig:Dectability}). 

Figure~\ref{fig:Confusion} shows that, whichever estimator is used, the expected confusion rates for detections in the nearby volume ($D< 40$ Mpc) are low and comparable to that for the ALFALFA and HIPASS surveys \citep{Jones2015}. The impact of this effect on the HIMF was studied by \cite{Jones2015}, who found that the bias introduced by confusion to the ALFALFA and HIPASS HIMF Schechter fit parameters was within the published random errors, with a 1--2$\sigma$ suppression for the faint-end slope $\alpha$. This suggests that the impact of confusion on the science goal of constraining the low-mass end of the HIMF with CHORD (Section~\ref{sec:HIMF_Constrain}) will be similarly small. 

In contrast, we find that the fraction of confused detections beyond $z > 0.1$ rises to $\gtrsim 55\%$, as shown by blue solid line in the left panel of Figure~\ref{fig:Confusion}. However, the total mass contribution by all sources in the beam ($M_{\rm{stack}}$) is on average $\lesssim 20\%$ of the \HI-mass detection limit ($M_{\rm{lim}}$) beyond a redshift of $z > 0.2$, as shown by the black dash-dotted line. The scatter in $M_{\rm{stack}}/M_{\rm{lim}}$ reduces significantly at high redshifts since detections will be massive in comparison to the confusing sources in the beam, as shown by the gray shaded region. Thus the scenario in which blends of two or more non-detections exceed the detection threshold is highly unlikely beyond $z > 0.2$.

Figure~\ref{fig:Confusion} therefore implies that additional care will be required when constraining the high-mass slope of the HIMF at high redshifts, but confusion should not be a dominant effect. While it is improbable ($\lesssim 5\%$ chance) for two or more massive detections $> 10^{10} \solM$ to coincide spectrally and spatially beyond $z > 0.2$, the non-negligible contribution from the stack of non-detections will need to be accounted for. In this respect, CHORD's primary utility for high-$z$ \HI-galaxy science will be discovering rare massive galaxies in previously unexplored volumes. Other tracking interferometers with higher spatial and spectral resolution such as the Very Large Array (VLA) can then follow up CHORD's findings for more accurate estimates of source mass and flux. 

Considering the significant source counts in the range at which confusion and RFI contamination from GNSS satellites is expected to be significant, we revisit the source counts from the HIMF predictions shown in Figure~\ref{fig:HIMF_constrain}. If we conservatively disregard all detections in the $D \gtrsim 40$ Mpc and $z \lesssim 0.3$ distance range, a 5-year CHORD survey is still expected to detect on the order of $10^4$ sources with $D \lesssim 40$ Mpc and $z \gtrsim 0.3$. This conservative source count estimate is comparable to the number of detections from the ALFALFA $\alpha.100$ catalog and the first FASHI data release. However, several of these $10^4$ CHORD detections (predicted to be unaffected by confusion and RFI) are expected to be new discoveries, lying outside the $10^7 < \MHI/\solM < 10^{10}$ mass range for which the HIMF is already robustly measured by previous surveys. Confusion at these nearby and high redshift distances will therefore not significantly impact the CHORD \HI-science goals of constraining the low-mass and high-mass end of the HIMF. The impact of RFI and confusion is more significant for the 1-year survey, for which a higher fraction of anticipated detections fall in this range. The redshift reach of the CHORD 1-year survey will therefore be limited by RFI and confusion (see also Figure~\ref{fig:Highz}).

We note that the high confusion rate at intermediate distances in CHORD data presents an opportunity to connect disparate science cases. Studies such as \cite{Townsend2026} have shown that direct detections of galaxies can help validate intensity mapping measurements. Since CHORD will simultaneously conduct an \HI\ galaxy survey and carry out \HI\ intensity mapping at the same redshifts, the survey volume between $D \gtrsim 40$ Mpc and $z \lesssim 0.3$ can be leveraged as a semi-blended bridge between individual detections and a fully integrated signal for intensity mapping validation. Considering that the distribution of galaxies has been extensively mapped in this distance regime by optical surveys like SDSS and DESI \citep[e.g.,][]{SDSS_DR, DESI_DR}, CHORD will be able to compare this distribution to the \HI\ signal and provide a valuable consistency check for source clustering models \citep[e.g.,][]{Obuljen2019}. 

\section{Discussion}
\label{sec:Discussion}
The implications of CHORD survey forecasts are discussed in this section. A comparison to other ongoing and future \HI\ surveys is provided in Section~\ref{sec:SurveysCompare}. The impact of cosmic variance on constraining the HIMF with CHORD surveys is addressed Section~\ref{sec:CosmicVariance}, and in  Section~\ref{sec:DarkGalaxies}, we approximate the number of dark galaxy candidates that may be found with CHORD.

\subsection{Comparisons to other \HI\ Surveys}
\label{sec:SurveysCompare}
Here we place the CHORD survey in the larger context of ongoing and future untargeted \HI\ galaxy surveys leading up to the Square Kilometre Array (SKA) era of \HI-science. 

FASHI on FAST has surveyed a sky area of $\sim19500\ \rm{deg}^{2}$ from $-14^{\circ} < \delta < +66^{\circ}$, which partially overlaps with CHORD \citep{Zhang2021, Kang2022}. FASHI's first data release included 41,741 sources from which the HIMF was estimated down to $\MHI \sim 10^{7} \solM$ \citep{FASHI_DR, Ma2025}, and the final source catalog includes 156,411 sources from which an HIMF down to $\MHI \sim 10^{6.2} \solM$ was reported \citep{FASHI_DR2}. The median FASHI sensitivity of 0.57 mJy/beam per 6.4 \kms\ channel is similar to our forecasts for a 1-year CHORD survey; a 5-year CHORD survey would be twice as sensitive as FASHI (albeit over a smaller sky footprint). 

Additional \HI\ survey initiatives with FAST are also underway,  such as the FATHOMER \citep[FAst neuTral HydrOgen intensity Mapping ExpeRiment,][]{Li2023} pilot survey which covered 60 deg$^{2}$ on sky for a total observation of 28 hours. FATHOMER identified a total of 702 detections with $\MHI > 10^{6.2} \solM$ and $z < 0.09$, and 77 of them lack optical counterparts and are possible dark or optically-faint candidates \citep{Shu2026}. CHORD is expected to $\sim 30$ dark galaxies based on the number density estimates from \cite{Zheng2025} for sources with even lower masses ($10^{5} < \MHI/\solM < 10^{6}$), as discussed in Section~\ref{sec:DarkGalaxies}.

The Deep Synoptic Array \citep[DSA;][]{DSA2000} is currently under construction in Nevada, USA, and will provide an \HI\ census of the Northern hemisphere along with CHORD. The two telescopes are highly complementary: the DSA's 1650 $\sim$6-m dishes should have a similar point-source sensitivity to CHORD, but the DSA's dish distribution includes baselines up $\sim 15$ km, producing a synthesized beam FWHM of $\sim$3.5\arcsec\ in sharp contrast to CHORD's $\sim$5\arcmin\ synthesized beam. The DSA will perform a \HI\ extragalactic survey $\sim 31\,000$ deg$^{2}$ on sky over 5 years, and its design is ideal for detecting \HI\ in galaxies at $z\gtrsim 0.1$. In the local volume, the DSA will spatially resolve gas-rich galaxies with a column density sensitivity that will detect dense \HI\ clumps in star forming disks, but diffuse emission will be missed. The CHORD design will therefore be better suited to discovering diffuse dwarfs in the local universe, including optically faint and dark galaxies. 

There are also a number of untargetted \HI\ surveys in the Southern Hemisphere that are completed or currently underway. In the local universe, WALLABY \citep[The Widefield ASKAP L-band Legacy Allsky Blind surveY;][]{ Koribalski2020,Murugeshan24} on the Australian SKA Pathfinder (ASKAP) is forecasted to detect $2 \times 10^{5}$ sources out to $z = 0.1$ across its $\sim 14\,000$ deg$^{2}$ footprint, most of which will be spatially resolved by its 30\arcsec\ synthesized beam. MIGHTEE-HI \citep[MeerKAT International GigaHertz Tiered Extragalactic Exploration;][]{MIGHTEE2021} on MeerKAT \citep{Jonas2016} and DINGO \citep[Deep Investigation of Neutral Gas Origins;][]{Rhee23} on ASKAP covers $\sim 60$ deg$^{2}$ (at 1420 MHz) with sufficient sensitivity to detect \HI\ sources out to $z \sim 0.4$. LADUMA \citep[Looking at the Distant Universe with the MeerKAT Array;][]{Blyth16} on MeerKAT will go deeper still, aiming to detect individual systems at redshifts as high as $z \sim 1$ in a single $\sim 1$ deg$^{2}$ (at 1420 MHz) pointing. These designs and strengths of these different surveys are all complementary to CHORD.


\subsection{Cosmic Variance}
\label{sec:CosmicVariance}

\begin{figure}[tbh!]
    \centering
    \includegraphics[width=\linewidth]{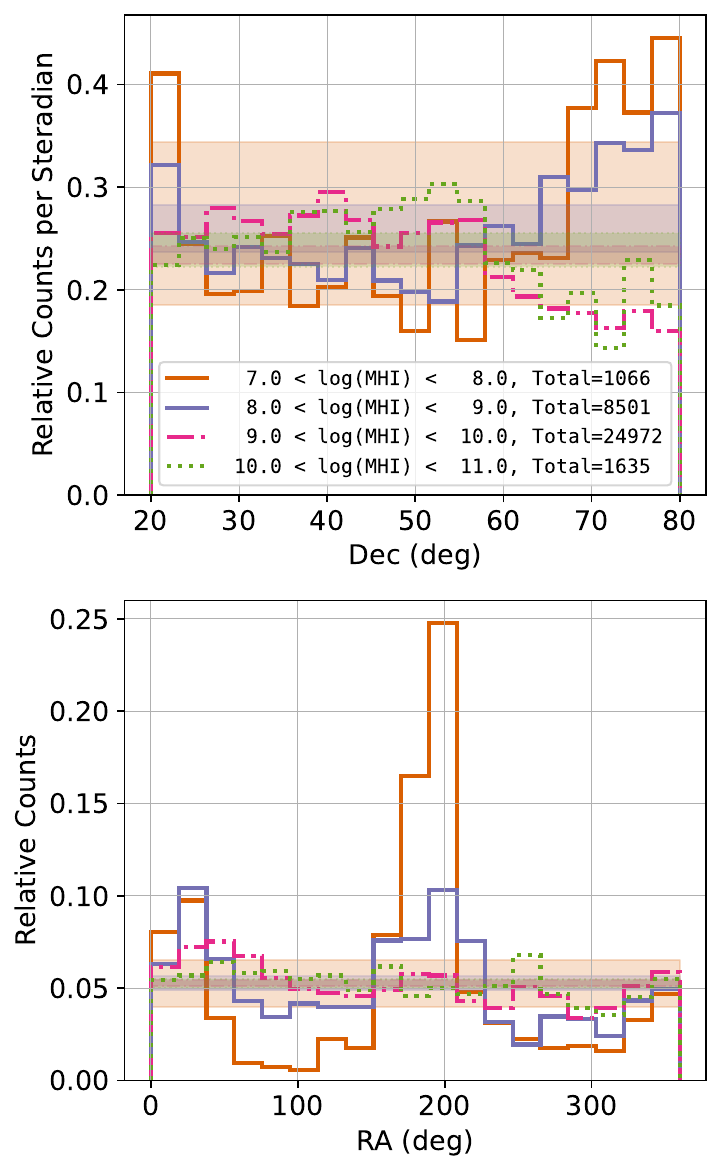}
    \caption{Spatial distribution of mock-\HI\ galaxies above the ALFALFA detection limit from \cite{Brooks2023}, extracted from the SIBELIUS-DARK constrained simulation \citep{SibeliusDark} over the declination range $+20^{\circ} < \delta < +80^{\circ}$ accessible to CHORD. The histograms shown the relative counts, where the counts in each mass bin has been normalized by the total counts in the mock-ALFALFA catalog. The top panel is additionally normalized by the solid angle covered by each declination bin to account for the changing sky area. The mean and standard deviations in the relative counts are shown by the shaded bands for each mass bin in matching colors for comparison. These values are estimated from our HIMF-drawn catalog with uniform spatial distribution for the same volume of \mbox{200 Mpc} as the constrained simulation, with the ALFALFA sensitivity cut applied. The RA and Dec bins are 18$^{\circ}$ and 2.4$^{\circ}$ wide, respectively. The large increase in relative counts seen around 180$^{\circ}$ in RA corresponds to the Virgo cluster.}
    \label{fig:CosmicVariance}
\end{figure}

In this section, we forecast the impact of cosmic variance on CHORD's HIMF constraints. Cosmic variance is defined here as a source of uncertainty arising from the large-scale structure of the Universe. While the Universe is homogeneous and isotropic on large scales, local density variations can bias mass function measurements if a finite survey volume does not adequately sample different  environments. To estimate the scale of this effect, we use a mock-ALFALFA catalog created by \cite{Brooks2023}, based on the \textsc{sibelius-dark} constrained N-body dark matter only simulation \citep{SibeliusDark}.

\textsc{sibelius-dark} is part of the `Simulations Beyond The Local Universe' \textsc{sibelius} project which aims to understand the Local Group in a cosmological context by embedding a Local Group analog into a primordial density field \citep{Sibelius2022}. \textsc{sibelius-dark} then uses a semi-analytic model of galaxy formation, \textsc{galform} to calculate the galaxy population \citep{GALFORM}. The resulting high-resolution large-scale structure is constrained out to a distance of \mbox{200 Mpc}, including the Virgo, Coma, and Perseus galaxy clusters \citep{SibeliusDark}. This volume is ideal for comparison with ALFALFA which observed galaxies out a distance $\sim 216$ Mpc. 
To generate a mock-ALFALFA survey, \cite{Brooks2023} add stellar and gas components to the \textsc{sibelius-dark} galaxy rotation curves, produce \HI\ spectra from the total gas estimates, and apply the ALFALFA detection limit. In their work, they additionally apply the ALFALFA survey boundaries, but have provided us the full all-sky mock-ALFALFA catalog to which we apply the CHORD survey boundaries. Figure \ref{fig:catalogs} shows their mock-ALFALFA catalog with large-scale structure for the CHORD footprint.

Figure \ref{fig:CosmicVariance} shows the normalized and sky-area corrected source counts from the mock-ALFALFA at different positions within the CHORD footprint. We find that the lowest ($10^7 < \MHI/\solM < 10^9$) mass bins show the most variation in normalized counts as a function of right ascension and declination. In contrast, the most populated mass bin $10^9 < \MHI/\solM < 10^{10}$ and the highest mass bin $10^{10} < \MHI/\solM < 10^{11}$ have roughly comparable number of detections everywhere in the sky, a difference that is explained by the larger survey volume probed at higher \HI\ masses \citep[c.f.\ Figure~\ref{fig:Dectability}; e.g.][]{Schneider2008}.

It is worth noting that this constrained simulation is not meant to predict the exact positions of \HI\ galaxies beyond the ALFALFA footprint, particularly at high declinations where fewer observations exist. Rather, this example serves to illustrate how local density fluctuations can bias measurements of the HIMF at the low-mass end specifically since the survey volume available to dwarfs will be systematically more shallow. This was seen in the ALFALFA survey for which \cite{ALFALFA_Jones2018} reported more than a 3-$\sigma$ difference in HIMF shapes for the `Spring' (northern Galactic hemisphere) and `Fall' (southern Galactic hemisphere) skies. This difference is likely driven by the presence of the Virgo cluster in the foreground of the Spring sky, resulting in a steeper HIMF low-mass slope than the Fall sky which contained a void at similar distances. The Virgo cluster is seen at an RA of $180^{\circ}$ in Figure \ref{fig:CosmicVariance}. 

We now discuss these findings in the context of CHORD. Our fiducial 1-year CHORD footprint would map $\sim 8700$ deg$^{2}$ on the sky, which is roughly twice as much as the ALFALFA Spring sky ($\sim$ 4060 deg$^{2}$) and more than three times larger than the Fall sky ($\sim$ 2500 deg$^{2}$). This will be further superseded by the 5-year CHORD survey which will cover 13,000 deg$^{2}$. The increased sky coverage will therefore sample a wider range of large-scale environments, reducing uncertainty from cosmic variance. Nonetheless, the uncertainties from cosmic variance will still be higher for low-mass galaxies. For instance, sources with $\MHI < 10^{7} \solM$ will only be visible to CHORD out to a distance of $\sim$ 40 Mpc and $\MHI < 10^{6} \solM$ sources within a distance of $\sim$ 10 Mpc (see Figure~\ref{fig:Dectability}). Thus the new CHORD constraints on the low-mass end of the HIMF may be susceptible to cosmic variance.  

\subsection{Searching for Dark Galaxies}
\label{sec:DarkGalaxies}
We have shown that CHORD will be sensitive to the \mbox{$10^{5} < \MHI/\solM < 10^{6}$} mass regime out to $\sim$ 10 Mpc for a 5-year survey, enabling the largest untargeted search for nearby dark galaxies to date. However, forecasting dark galaxy detection counts remains uncertain as the intrinsic number density of this population is not yet well constrained and depends heavily on the chosen implementation of baryonic physics in simulations \citep[e.g.][]{Benitez2017, Lee2024}. 

For instance, in the Hestia and Auriga simulations \cite{Zheng2025} find 89 HIDES with $M_{\star} \lesssim 10^{6}\,\solM$ and $\MHI > 10^{5} \,\solM$, only one of which is completely starless. They determine that the number density of HIDES within 3.7 Mpc of the Milky Way (MW) is best fit by \mbox{$n=0.25(d_{\mathrm{MW}}/\mathrm{Mpc})^{-1.4}$ Mpc$^{-3}$}. This number density approximates to $\sim 14$ HIDES in the Local Group. 

To estimate the total number of HIDES available to a 5-year CHORD survey, we count the number of MW analogs ($10^{9} < \MHI/\solM < 10^{10}$) within $5 < D/\rm{Mpc} < 10$, excluding sources located closer than 5 Mpc to avoid confusion with HVCs. We find 14 MW analogs in the HIMF-drawn uniformly distributed catalog and 12 in the mock-ALFALFA constrained simulation with large-scale structure. Assuming each MW analog has a similar environment to the Local Group with $\sim 14$ dark galaxies in its vicinity, we approximate that about 168--196 HIDES should be present in the CHORD $+20^{\circ}<\delta <+80^{\circ}$ sky. However, not all the HIDES available in the CHORD footprint will necessarily be detected. Within a distance range of $5 < D/\rm{Mpc} < 10$, CHORD is expected to detect roughly 15\% of the $10^{5} < \MHI/\solM < 10^{6}$ sources expected from the HIMF. Applying this 15\% detection rate to the HIDES, we approximate that CHORD will detect about $\sim 30$ HIDES-type dark galaxy candidates. These candidates can then be followed up with higher angular and spectral resolution instruments to characterize their morphology and spectral line widths to confirm or reject their candidate status.

In comparison, \cite{Garcia2026} use four realizations of zoom-in Hestia and Nivaria-LG Local Group simulations and predict that 8 dark galaxies can be observed within 2.5 Mpc of the Local Group using the FAST telescope, based on its sky coverage and sensitivity. This estimate is roughly consistent with the $\sim 14$ HIDES expected within 3.7 Mpc from \cite{Zheng2025}. In practice however, dark galaxy detections within 2.5 Mpc will be difficult to differentiate from HVCs. 

We note that since the \cite{Zheng2025} HIDES number density is for MW analog environments, this estimate does not account for isolated dark galaxies, which CHORD will additionally find. Isolated candidates are likely to be stronger dark galaxy candidates than HIDE analogs or Cloud-9 as the \HI-gas is more likely to be of cosmic origin belonging to a dark matter halo than a stripped gas cloud from tidal or ram pressure interactions \citep{Benitez2023, Benitez2024}.  

\section{Conclusion}
\label{sec:Conclusion}

CHORD will be the first highly-redundant interferometer to carry out an extragalactic \HI-survey. We use HIMF-drawn mock catalogs with spectral profiles described by the busy function to forecast untargeted \HI-survey outcomes with CHORD. 

We find that a 1-year CHORD survey can achieve a \mbox{0.6 mJy} sensitivity for 5 \kms\ spectral resolution, detecting $\sim 3\times 10^{5}$ sources within a 8700 deg$^{2}$ footprint out to a distance of $z \sim 0.3$. A 5-year survey covering \mbox{13,000 deg$^{2}$} can detect $\sim 10^6$ sources out to $z \sim 0.5$ with a sensitivity of 0.3 mJy for 5 \kms\ resolution. Accounting for confusion and RFI, a 5-year CHORD survey will expand the sample of \HI-selected observations, with thousands of discoveries of the currently undersampled gas-rich low-mass and high-mass populations. We summarize our main forecasts with respect to different distance regimes, as follows. 

In the nearby universe, if the intrinsic number densities of gas-rich dwarfs can be extrapolated from the HIMF, CHORD will find thousands of $\MHI < 10^{7} \solM$ sources out to a distance of $\sim 40$ Mpc, and dozens of $\MHI < 10^{6} \solM$ sources out to $\sim 10$ Mpc. This will place new constraints on the low-mass end of the HIMF down to $\sim 10^{5.5} \solM$. Cosmic variance will be a source of uncertainty for these constraints because the low-mass sources will be biased to the large-scale structure of the local universe. Spectroscopic confusion will be negligible for nearby sources, with blends expected to occur less than 10\% of the time. Based on number density estimates of \HI-rich dark galaxies in Local Group simulations from \cite{Zheng2025}, we expect to find $\sim 30$ HIDES-type dark galaxies in a 5-year CHORD survey. The low-mass population census provided by CHORD will therefore place observational constraints on the mass threshold of galaxy formation.

At intermediate distance scales ($D > 40$ Mpc and $z < 0.3$), a 5-year CHORD survey is expected to detect on the order of $10^{6}$ galaxies. However the detections are expected to be confused about 60\% of the time with a source that are least 10\% of the detection mass ($\geq 0.1M_{\rm{det}}$), and about 25\% of the time with other detections. This regime can be leveraged as a semi-blended bridge between \HI-galaxy survey and \HI-intensity mapping science, allowing for consistency checks and validation of intensity mapping measurements. The redshift range $0.1 < z < 0.3$ (1090--1290 MHz) is also expected to be severely contaminated by RFI from GNSS satellites. However, there is ongoing investigation within the CHORD collaboration to characterize whether this RFI can be mitigated by leveraging CHORD millisecond time resolution data. 

At high redshifts ($0.3 < z < 0.5$), CHORD is expected to detect on the order of $10^3$ sources with $\MHI > 10^{10.5} \solM$. These detections will be confused with $\geq 0.1M_{\rm{det}}$ sources roughly 60\% of the time. However, the combined contribution from all stacked sources in the beam will be only about 20\% of the detection mass limit. Furthermore, confusion with other massive detections will be unlikely, occurring less than 3\% of the time. This will allow CHORD to discover rare giant galaxies at high redshift, albeit with a $\sim 20\%$ uncertainty in their mass measurements. The detections can then be followed up with deeper targeted observations for more accurate mass and flux measurements. CHORD will thus deliver a census of high-mass galaxy detections over an unprecedented redshift range, placing new constraints on the high-mass slope of the HIMF and its possible redshift evolution.

While the focus of this work has been forecasting HIMF constraints from CHORD \HI\ surveys, there are many other \HI\ science cases that CHORD will be well suited to explore. For instance, the higher sensitivity of CHORD may likewise improve constraints on the \HI\ width function (HIWF) which is a sensitive tracer of the dark halo mass function \citep[e.g.,][]{Moorman2014, Oman2022, Brooks2023}. The CHORD \HI\ data can also be combined with complimentary surveys to improve measurements of galaxy scaling relations such as the Baryonic Tully Fisher Relation (BTFR) as was previously done for ALFALFA by \cite{Ball2023}. Additionally, different techniques can be employed to detect sources beyond the sensitivity and volume limits of CHORD, such as spectral stacking analyses to extract the statistical \HI\ properties of galaxies below the detection limit \citep[e.g.,][]{Fabello2012, Delhaize2013, Brown2015, Healy2019} and gravitational lensing to detect sources beyond the survey volume \citep[e.g.,][]{Hunt2016, Blecher2019, Chakraborty2023}. The expanded parameter space of \HI\ sources that will be probed by CHORD will thus provide new observational constraints on models of galaxy formation and evolution.  

\section*{Acknowledgments}
AL acknowledges support from the Natural Sciences and Engineering Research Council of Canada through their Discovery Grants Program and their Alliance International Program, as well as the William Dawson Scholar program at McGill University.

\bibliography{references}{}
\bibliographystyle{aasjournal}


\end{document}